\documentclass[conference,compsoc]{IEEEtran}
\IEEEoverridecommandlockouts

\usepackage[T1]{fontenc}
\usepackage{amsmath,amssymb,amsthm}
\usepackage{booktabs}
\usepackage{cite}
\usepackage[hidelinks]{hyperref}
\usepackage{tikz}
\usetikzlibrary{positioning, fit, backgrounds, calc}
\usepackage{xspace}
\usepackage{pifont}
\usepackage{url}
\usepackage{enumitem}
\usepackage{graphicx}
\usepackage{multirow}
\usepackage[ruled,vlined,linesnumbered]{algorithm2e}

\usepackage{xcolor}
\usepackage[normalem]{ulem}
\usepackage{hyperref}

\definecolor{linkblue}{HTML}{5AA9E6}

\SetAlgoCaptionSeparator{:}
\SetAlCapNameFnt{\small\bfseries}
\SetAlCapFnt{\small\bfseries}
\SetAlgoNlRelativeSize{-1}
\SetKwInput{KwIn}{Input}
\SetKwInput{KwOut}{Output}

\makeatletter
\renewcommand{\algocf@captiontext}[2]{%
  \small\bfseries #1 #2%
}
\makeatother

\usepackage{float}
\usepackage{tabularx}
\DontPrintSemicolon
\SetKwInput{KwIn}{Input}
\SetKwInput{KwOut}{Output}
\graphicspath{{figures/}}

\newif\ifshowcomments
\showcommentsfalse

\ifshowcomments
  \newcommand{\CommentA}[2]{\textit{\textbf{\textcolor{red!65!black}{#1}}: \textcolor{gray!65!black}{#2}}}
  \newcommand{\ZM}[1]{\CommentA{JZM}{\textcolor{red}{#1}}}

  \newcommand{\CommentB}[2]{\textit{\textbf{\textcolor{blue!65!black}{#1}}: \textcolor{gray!65!black}{#2}}}
  \newcommand{\ZZZCCC}[1]{\CommentB{ZC}{\textcolor{blue}{#1}}}
\else
  \newcommand{\ZM}[1]{}
  \newcommand{\ZZZCCC}[1]{}
\fi

\newcommand{\pspace}[0]{2pt}

\theoremstyle{definition}

\begin{document}
\newcommand{\sysname}{\textsc{DCIGSE}\xspace}
\title{Breaking the Coupling: Efficient Concurrency Fuzzing for Linux Kernel Data Race Detection}

\author{
\IEEEauthorblockN{Cen Zhang}
\IEEEauthorblockA{
\textit{Beihang University}\\
Beijing, China\\
\textit{zzzccc427@buaa.edu.cn}
}
\and
\IEEEauthorblockN{Jia-Ju Bai\textsuperscript{*}%
\thanks{\textsuperscript{*}Corresponding author.}}
\IEEEauthorblockA{
\textit{Beihang University}\\
Beijing, China\\
\textit{baijiaju@buaa.edu.cn}
}
\and
\IEEEauthorblockN{Zi-Yu Zhang}
\IEEEauthorblockA{
\textit{Beihang University}\\
Beijing, China\\
\textit{ziyuzhang201@buaa.edu.cn}
}
\and
\IEEEauthorblockN{Zuming Jiang}
\IEEEauthorblockA{
\textit{The University of Hong Kong}\\
Hong Kong SAR, China\\
\textit{jzuming@hku.hk}
}
}


\maketitle

\begin{abstract}
Detecting data races in the Linux kernel requires exploring both the input space and the thread-interleaving space. Existing kernel concurrency fuzzers couple these two searches by exploring interleavings during fuzzing, often spending substantial scheduling effort on inputs with little potential to trigger real races. In this paper, we present a new concurrency fuzzing paradigm that decouples concurrent-input generation from thread-interleaving exploration, enabling expensive scheduling to focus on promising inputs.

To identify inputs that warrant targeted scheduling, we introduce May Race Pairs (MRPs), which represent conflicting memory-access pairs observed within a dynamically determined time threshold. Our framework first generates and executes concurrent syscall groups without controlled scheduling, collecting MRPs to guide further input mutation and identify schedule-worthy inputs. It then uses a separate scheduling instance to explore targeted interleavings and confirm real data races. To improve the effectiveness of both stages, we incorporate LLM-assisted semantic mutation for generating diverse syscall groups, backpressure-guided dynamic threshold control for balancing candidate production and scheduling consumption, and history-aware context recovery, context-aware interleaving scheduling, and candidate MRP prioritization for efficient race confirmation.

We implement our framework for the Linux kernel and evaluate it on eight common kernel modules in Linux 6.17. It discovers 187 unique real-world data races, including 59 harmful races. Under the same testing budget, our approach detects substantially more data races than existing kernel concurrency fuzzers, including SegFuzz and CONZZER. These results demonstrate the effectiveness of decoupling concurrent-input generation from interleaving exploration and concentrating scheduling effort on promising candidates.

\end{abstract}

\section{Introduction}
\label{sec:intro}
Data races are a common and critical class of concurrency issues in the Linux kernel.
A data race occurs when a conflicting access pair executes concurrently. A conflicting access pair is defined as two accesses from different threads to the same or overlapping memory range, where at least one access is a write.
In the Linux kernel, data races can cause data corruption~\cite{cve-2022-49171}, memory corruption~\cite{cve-2022-49554,cve-2022-50145,cve-2025-40318}, and severe security vulnerabilities such as privilege escalation~\cite{cve-2021-4202,cve-2023-0266,dirtycow}.


Detecting data races is more challenging than detecting single-threaded bugs, because race exposure depends on both
suitable inputs and schedules.
The input dimension determines whether concurrently executed threads can create a conflicting access pair. The schedule dimension determines whether these accesses occur simultaneously. Therefore, detecting data races in the Linux kernel must consider both input and schedule dimensions. This greatly enlarges the search space and complicates practical race detection.


Many approaches have been proposed to detect data races in the Linux kernel.
Coverage-guided fuzzers primarily explored the input dimension alone~\cite{angora,fairfuzz,smartgreybox,superion,syzkaller,moonshine,difuze,charm,janus,hydra}, relying on runtime data-race checkers such as KCSAN~\cite{kcsan} and KTSAN~\cite{ktsan} to identify races during execution.
However, their main feedback signal is code coverage, which reflects sequential system executions but provides little guidance for exploring thread interleavings, and therefore limits their ability to expose data races.


Recent work has increasingly recognized that data-race detection must reason about both inputs and thread interleavings. One line of work first reduces the search space through static race analysis, memory-access traces collected from sequential executions, or learned predictors trained from prior executions, and then performs concurrent execution with strategic scheduling within the reduced space~\cite{razzer,snowboard,snowcat}. Another line of concurrency-aware fuzzing uses runtime concurrency information to guide thread-interleaving exploration, increasing the probability of triggering concurrency issues~\cite{muzz,confuzz,liu2018heuristic,conzzer,segfuzz,sect}. These approaches substantially improve practical race detection because they move beyond single-threaded coverage and consider schedule dimensions.


However, existing approaches remain inefficient because they still deeply couple their concurrent-input generation with thread-interleaving exploration when detecting data races. 
Specifically, existing approaches have to repeatedly execute the inputs and insert scheduling points (\emph{e.g.}, thread delay) during their executions.  However, most of the inputs they randomly generate have limited potential to trigger real races, as a result of which existing approaches waste a significant amount of computing resources in scheduling these inputs. The root cause of this inefficiency is that concurrent-input generation and thread-interleaving exploration remain coupled. 
As they lack a systematic way to identify the inputs that have a higher potential to trigger data races under scheduling,
existing approaches either have to schedule most inputs once they are generated, including many low-potential ones, or rely on concurrency feedback, such as covered thread interleavings, whose collection requires repeatedly scheduling for each generated input.

In this paper, we propose a new concurrency fuzzing paradigm, \textbf{\textit{decoupling concurrent-input generation and thread-interleaving exploration}}, to significantly increase the efficiency of race detection in the Linux kernel. Controlled scheduling is much more expensive than ordinary concurrent execution, as it typically requires repeated input executions or fine-grained system execution controls. Therefore, our insight is that the fuzzers should perform thread scheduling for only the inputs that have a higher potential to trigger data races, instead of applying expensive scheduling to every generated input.
Based on this insight, we separate data race detection into two decoupled processes: (1) generating concurrent inputs and selecting the ones that deserve to be scheduled; (2) scheduling the execution of the given input and determining whether it can indeed trigger a data race.





The core part of such decoupling is how to determine whether a given input has the potential to trigger real races and thus deserves the expensive scheduling. To address it, we introduce a new concept, \textbf{\textit{May Race Pair}} (MRP). 
An MRP represents a conflicting access pair that occurs within a dynamically determined time threshold. It does not prove the existence of a data race. Instead, it indicates a scheduling-worthy pair: when two accesses target the same memory address and occur within a short time window, they are more likely to occur concurrently under scheduling than other pairs. With MRP, we can effectively select the inputs that have a higher potential to trigger data races and perform expensive scheduling on only these inputs. Moreover, an MRP records the conflicting access pair, which enables the scheduling for the input to focus on a pair of concrete code sites, significantly reducing the scheduling cost.




Based on this concept, we design \sysname, a kernel concurrency fuzzing framework. \sysname consists of two decoupled instances: an input-generation instance and a thread-scheduling instance. The two instances run in parallel. The input-generation instance produces inputs that trigger MRPs, while the thread-scheduling instance consumes these inputs by scheduling their thread execution to trigger the potential data races. 
To maximize the number of detected data races, the throughput of these instances should be the same, which motivates our design of the dynamically determined time threshold specified in the MRP.
This threshold involves a tradeoff: (1) a smaller threshold admits only conflicting access pairs with shorter temporal distances, which are usually more promising for triggering real races under scheduling, but may produce too few candidate MRPs and underutilize the thread-scheduling instance; (2) a larger threshold allows the input-generation instance to collect more MRPs, but may also admit more low-potential candidates, which may waste the scheduling effort. 
\sysname adopts a backpressure-guided control to dynamically adjust the threshold. Specifically, \sysname decreases the time threshold when the collected MRPs accumulate faster than they can be consumed, so that the input-generation instance admits fewer MRPs with shorter temporal distances and the thread-scheduling instance can spend its limited budget on more promising candidates. Conversely, \sysname increases the time threshold when the collected MRPs are insufficient, ensuring enough candidate MRPs to avoid starvation of the thread-scheduling instance.

The input-generation instance generates syscall groups. Each group consists of two sequences of syscalls. Given a syscall group, this instance executes the two syscall sequences concurrently without controlled scheduling. The instance collects the syscall groups that trigger new MRPs. We propose LLM-assisted semantic mutation for generating more interesting syscall groups. Specifically, we provide the LLM with module-specific mutation constraints and cross-sequence semantic interaction objectives. Guided by this information, we leverage its reasoning capabilities and pretrained knowledge to understand and semantically mutate promising syscall groups selected based on MRP feedback.

The thread-scheduling instance consumes the collected syscall groups and schedules them to confirm whether they can indeed trigger data races. We integrate three methods: history-aware context recovery to recover the kernel state required to reproduce candidate MRPs, context-aware interleaving scheduling to distinguish different dynamic occurrences of the same static access site using finer-grained runtime contexts, and candidate MRP prioritizing to prioritize candidate inputs based on previous scheduling results.

We implemented \sysname and evaluated \sysname on eight common Linux kernel modules in Linux 6.17. In total, it found 187 unique real-world data races, among them 59 are identified as harmful races. We reported the harmful races. As of paper submission, 39 bugs have been confirmed, and 15 of the bugs have been fixed. These races may result in critical security vulnerabilities like concurrent use-after-frees.
The comparison results indicate that our approach can find significantly more data races than existing approaches including SegFuzz~\cite{segfuzz} and CONZZER~\cite{conzzer} within the same testing budget.


In summary, we make the following contributions:

\begin{itemize}[leftmargin=*, itemsep=0.2em, topsep=0.2em]

\item At the conceptual level, we propose a new paradigm for kernel concurrency fuzzing: decoupling concurrent-input generation and thread-interleaving exploration. We further introduce a new concept, May Race Pair (MRP), to facilitate the decoupling.


\item At the technical level, we design \sysname, a kernel concurrency fuzzing framework that realizes this paradigm. We integrate \sysname with LLM-assisted semantic mutation to generate diverse system inputs, and with history-aware context recovery, context-aware interleaving scheduling, and candidate MRP prioritizing to efficiently explore promising thread interleavings.


\item At the practical level, we implemented and evaluated \sysname on eight common Linux kernel modules in Linux 6.17. \sysname found 187 unique real-world data races (59 harmful, 39 confirmed, and 15 fixed), outperforming the existing approaches. 

\end{itemize}

\section{Background}
\label{sec:bg}

\subsection{Data Race}


In this work, consistent with prior kernel race fuzzers (e.g., Razzer and CONZZER~\cite{razzer,conzzer}), we identify a data race when the two endpoints of a conflicting access pair are observed to overlap in execution. A conflicting access pair consists of two accesses from different threads to the same or overlapping memory range, with at least one access being a write. Thus, triggering a data race requires an input that produces a concrete conflicting access pair and a schedule that brings its two accesses into monitored overlap in time.

\subsection{Concurrent Execution in the Linux Kernel}

System calls (syscalls) are the primary interface through which user-space programs drive kernel behavior. A syscall can create and manipulate kernel objects, establish resource dependencies, update subsystem state, and trigger different allocation, access, and cleanup paths. In kernel fuzzing, syscalls serve as the de facto input representation.


To systematically exercise interactions between different threads, a concurrency fuzzer executes multiple system-call sequences concurrently. These executions may interact through the same object instance, related resources, or shared subsystem state, and may consequently produce a concrete conflicting access pair.

These concurrent executions introduce a large interleaving space. Different syscall sequences may reach different shared states and trigger diverse memory accesses. Moreover, even the same concurrently executed sequences can interleave in many ways because each syscall handler contains numerous internal operations, blocking points, and memory accesses. To prevent interleavings from corrupting shared kernel data, the kernel relies on synchronization mechanisms to enforce safe thread execution orders. However, missing or incorrectly scoped synchronization may allow conflicting accesses to occur concurrently, manifesting as data races.

Accordingly, kernel data-race detection involves both an input dimension and a schedule dimension. The input dimension determines whether the executed syscall sequences produce a concrete conflicting access pair, while the schedule dimension determines whether the two accesses temporally overlap. 
Without proper inputs to establish the required kernel states and execution contexts, scheduling cannot drive the thread execution to reach potentially conflicting code paths. Conversely, inputs alone cannot ensure that these accesses actually overlap in time during execution.


\subsection{Limitations of Existing Approaches}

Existing approaches do not fully decouple concurrent-input generation and interleaving exploration. One line of work searches both spaces jointly in a single fuzzing process, while another first uses static analysis or observations from single-threaded executions to guide a later concurrent-testing stage, where inputs and schedules remain coupled.

KRACE and CONZZER follow the first design. KRACE jointly evolves multithreaded syscall programs and delay choices, while CONZZER jointly mutates program inputs and concurrent-call-pair targets~\cite{krace,conzzer}. Although KRACE runs its race checker in the background, this separates trace collection from race analysis. It does not decouple concurrent-input generation and interleaving exploration.

Snowboard, Razzer, SegFuzz, and Ozz instead use static analysis or runtime observations from single-threaded executions to select candidates and guide subsequent concurrent testing, but construct concurrent inputs and schedules together in that later stage~\cite{razzer,snowboard,segfuzz,ozz}. Snowboard illustrates this limitation. It records memory-access traces from independent sequential tests and infers cross-trace PMCs. Only afterward does it combine test pairs and schedule their concurrent execution. Snowboard reports that only about 36\% of the resulting tests exercise the predicted channel because concurrency may change runtime objects or control flow~\cite{snowboard}. Static or sequential information cannot accurately predict complex concurrent behavior, so whether a constructed concurrent input realizes the intended interaction remains unknown until the input is executed concurrently. Consequently, concurrent-input generation and interleaving exploration remain coupled in the later stage.

In contrast, \sysname decouples concurrent-input generation and interleaving exploration. Rather than relying on PMCs or other signals inferred from static analysis or single-threaded executions, \sysname uses MRPs collected while syscall groups execute concurrently. Each MRP both guides further mutation of its group and records a concrete conflicting access pair for subsequent scheduling. This allows concurrent-input generation to pursue diverse concurrent behaviors, while expensive scheduling is reserved for selected inputs and targets their recorded access pairs.

\section{Basic Idea}

\subsection{May Race Pair}
\label{sec:mrp}

We decouple concurrent-input generation and thread-interleaving exploration by selecting the inputs that have the potential to trigger data races and performing thread scheduling for only the selected inputs. In this case, this separation significantly reduces thread-scheduling cost and improves overall race-detection efficiency. The research question in this design is: how to determine whether a given input deserves to be scheduled?



Kernel executions produce many cross-thread accesses to overlapping addresses, but not all of them indicate a meaningful interaction between two threads. For example, two accesses may overlap only because the same address has been released and later reused for a different kernel object, rather than because both accesses operate on the same live kernel object. Moreover, even if the two accesses target the same kernel object, they may belong to entirely different lifecycle phases of the object, such as resource initialization and finalization, and thus are temporally isolated by a substantial time gap under normal execution.
Therefore, we use temporal closeness as an additional admission criterion for conflicting access pairs: an observed conflicting access pair is interesting only when the two accesses occur within a bounded time window during execution.

We propose a new concept, May Race Pair (MRP), which serves as the criterion for selecting interesting syscall groups. An MRP is an observed conflicting access pair whose two accesses occur within a short time threshold. 
Section~\ref{sec:overview-threshold} describes how this threshold is dynamically determined.
An MRP does not prove a race. Instead, it marks a memory conflict whose temporally close endpoints warrant targeted interleaving exploration.
During the concurrent execution of a syscall group, we record the runtime memory accesses. Each access is represented as
\[
    a=\langle n,s,\mathit{tid},\mathit{addr},\mathit{size},\mathit{type},\mathit{ts}\rangle,
\]
 Here, \(n\) is a compile-time 64-bit identifier derived from the instrumented access’s source/IR location and kind, while \(s\) is a 64-bit hash of its runtime kernel call chain, \(\mathit{tid}\) identifies the thread that performs the access, \(\mathit{addr}\) and \(\mathit{size}\) describe the accessed memory range, \(\mathit{type}\) indicates whether the access is a read or a write, and \(\mathit{ts}\) records the access timestamp.
Given two runtime accesses \(a_1\) and \(a_2\), we formalize a May Race Pair as
\[
\begin{aligned}
  & p = \langle a_1, a_2, \Delta t \rangle, \quad \Delta t = |a_1.\mathit{ts} - a_2.\mathit{ts}|, \\
  & \text{s.t.} \quad a_1.\mathit{tid} \neq a_2.\mathit{tid}, \\
  & \hphantom{\text{s.t.}} \quad [a_1.\mathit{addr}, a_1.\mathit{addr} + a_1.\mathit{size}) \\
  & \hphantom{\text{s.t.}} \quad \quad \quad \cap [a_2.\mathit{addr}, a_2.\mathit{addr} + a_2.\mathit{size}) \neq \emptyset, \\
  & \hphantom{\text{s.t.}} \quad a_1.\mathit{type} = \mathsf{W} \vee a_2.\mathit{type} = \mathsf{W}, \\
  & \hphantom{\text{s.t.}} \quad \Delta t \leq \tau.
\end{aligned}
\]

The first three constraints require the two accesses to form a conflicting access pair: they are performed by different threads, their accessed memory ranges overlap, and at least one access is a write. The last constraint is the temporal-admission constraint: the pair is retained as an MRP only when its temporal distance is no larger than the current threshold \(\tau\). The threshold \(\tau\) is an admission-control parameter rather than a race-confirmation criterion; Section~\ref{sec:overview-threshold} describes how \sysname adjusts it dynamically. For feedback deduplication, \sysname groups MRPs by their access-site pair \(\langle n_1,n_2\rangle\). MRPs with the same access-site pair but different stack-context pairs \(\langle s_1,s_2\rangle\) are treated as distinct context variants of that pair. For each access-site pair, \sysname retains at most a configurable number of context variants throughout the fuzzing campaign.

\subsection{Fuzzing Framework}
\label{sec:fuzzing-framework}

Based on MRPs, we propose a novel kernel concurrency fuzzing framework, \sysname, which decouples concurrent-input generation and thread-interleaving exploration using two instances.
Starting from a corpus of initial inputs, the input-generation instance generates and mutates inputs and executes them concurrently. This instance collects the new May Race Pairs from these executions.
The corresponding inputs of the new MRPs are stored in the corpus for further mutation, and also in the schedule-worthy input queue for subsequent scheduling.
The inputs in the schedule-worthy input queue are then consumed by the thread-scheduling instance.
Given an input, the thread-scheduling instance schedules its thread executions to force the conflicting access pair recorded in the corresponding MRP to occur concurrently. The success of such scheduling indicates that a real data race has been detected. The scheduled syscall groups are dequeued from the schedule-worthy input queue. In this framework, the two instances are connected only through schedule-worthy inputs.



\subsection{Dynamic Time-Threshold Control}
\label{sec:overview-threshold}

A conflicting pair is admitted as an MRP only if its temporal distance is at most \(\tau\), the admission-control parameter for the schedule-worthy corpus.
\(\tau\) trades candidate supply against selectivity. A smaller $\tau$ keeps only the inputs whose access pairs occur very close in time, making the retained inputs more selective and usually more promising for triggering data races under scheduling. However, an overly small threshold may produce too few schedule-worthy inputs and leave the scheduling instance underutilized. In contrast, a larger $\tau$ increases the candidate supply, but also admits inputs whose access pairs have larger temporal distances and are therefore harder to align under scheduling. Such pairs are less likely to be brought into overlap in time through scheduling and are more likely to include address-reuse artifacts, in which the same memory address is accessed after the underlying object has changed.


A fixed time threshold is unsuitable for our design. A suitable threshold varies across modules and, for the same module, different hardware configurations and workloads. It can also drift as the kernel state and the mix of syscall groups change over time.
More critically, the fixed threshold cannot 
maintain this balance across different kernel modules or throughout a fuzzing campaign.
The production speed of schedule-worthy inputs 
may change as concurrent-input generation reaches different subsystem states, while the consumption speed may also vary when scheduling different inputs. Consequently, the fixed threshold may either starve the scheduling instance or cause excessive candidate accumulation in the schedule-worthy input queue.

To tackle this problem, we propose backpressure-guided dynamic threshold control, as shown in Algorithm~\ref{alg:dynamic-threshold-control}. The
design is inspired by adaptive admission control and backpressure in staged event-driven service systems, where requests flow through explicit queues between stages and upstream stages react to downstream overload signals~\cite{welsh2003adaptive}. Similarly, we model the input-generation instance as a producer, the thread-scheduling instance as a consumer, and the schedule-worthy input corpus as the queue connecting them. The threshold $\tau$ controls admission into this queue: increasing $\tau$ supplies more schedule-worthy inputs, whereas decreasing it admits fewer but more promising candidates.

At the end of each control interval $T_c$, the controller measures the number of newly retained schedule-worthy inputs $P_k$, scheduled inputs $C_k$, and pending inputs $Q_k$. To reduce the effect of short-term fluctuations, it maintains exponentially weighted moving averages:
$$
\bar{P}_k
=
\rho\bar{P}_{k-1}+(1-\rho)P_k,
\qquad
\bar{C}_k
=
\rho\bar{C}_{k-1}+(1-\rho)C_k.
$$
where $\rho \in [0,1)$ is the smoothing factor. The controller estimates the buffered scheduling workload as
$$
W_k
=
\frac{Q_k}{\max\left(\bar{C}_k,\epsilon\right)}.
$$
where $\epsilon>0$ prevents division by zero. In this paper, we set
$T_c=30$s, $W_{\mathrm{low}}=10$, $W_{\mathrm{high}}=40$,
$\rho=0.8$, $\epsilon=1$, and use
$\gamma_{\mathrm{shrink}}=0.5$ and
$\Delta\tau=0.05(\tau_{\max}-\tau_{\min})$ as the multiplicative
tightening factor and additive relaxation step.

The controller maintains $W_k$ between a low and a high watermark. When the workload is excessive and continues to grow, it multiplicatively decreases $\tau$ to rapidly tighten admission. When the corpus is empty, or the workload is low and continues to decrease, it additively increases $\tau$ to gradually admit more candidates. Otherwise, the threshold remains unchanged during the current control interval. The two watermarks avoid frequent oscillations, while the asymmetric adjustment limits candidate accumulation without starving the thread-scheduling instance of candidate inputs.

\begin{algorithm}[t]
\small
\caption{Backpressure-Guided Dynamic Threshold Control}
\label{alg:dynamic-threshold-control}
\KwIn{initial threshold $\tau_0$, threshold bounds
$\tau_{\min}$ and $\tau_{\max}$, control interval $T_c$,
workload watermarks $W_{\mathit{low}}$ and $W_{\mathit{high}}$,
smoothing factor $\rho$, relaxation step $\Delta_\tau$,
tightening factor $\gamma_{\mathit{shrink}}$}
\KwOut{dynamically adjusted threshold $\tau$}

$\tau \leftarrow \tau_0$;
$\bar{P} \leftarrow 0$;
$\bar{C} \leftarrow 0$;

\While{\textsc{FuzzingRunning}()}{
\textsc{Wait}$(T_c)$;

$P \leftarrow \textsc{NewMRPRecords}()$\;
$C \leftarrow \textsc{ConsumedMRPRecords}()$\;
$Q \leftarrow \textsc{PendingMRPRecords}()$\;

$\bar{P} \leftarrow \rho\bar{P} + (1-\rho)P$\;
$\bar{C} \leftarrow \rho\bar{C} + (1-\rho)C$\;

$W \leftarrow Q / \max(\bar{C},\epsilon)$\;

\If{$W > W_{\mathit{high}}
    \land \bar{P} \geq \bar{C}$}{
    $\tau \leftarrow
    \max(
    \tau_{\min},
    \gamma_{\mathit{shrink}}\tau
    )$\;
}
\ElseIf{$Q = 0
    \lor
    (W < W_{\mathit{low}}
    \land \bar{P} \leq \bar{C})$}{
    $\tau \leftarrow
    \min(
    \tau_{\max},
    \tau+\Delta_\tau
    )$\;
}

\textsc{SetThreshold}$(\tau)$\;

}
\Return{$\tau$};
\end{algorithm}

\subsection{Concurrent Execution Model}
\label{sec:execution-model}

We use a syscall group as the input. A group is defined as the parallel composition of multiple syscall sequences, which are expected to execute concurrently:
\[
G = S_0 \parallel S_1 \parallel \cdots \parallel S_{m-1},
\]
where \(\parallel\) denotes parallel composition, and each \(S_i\) is assigned to a distinct execution context. Each syscall sequence is a list of syscalls that execute on the system sequentially:
\[
S_i=\langle c_{i,0},c_{i,1},\ldots,c_{i,n_i-1}\rangle ,
\]
where \(c_{i,j}\) denotes the \(j\)-th syscall in \(S_i\).



Because pairwise thread testing substantially reduces testing complexity while preserving most bug-exposing capability~\cite{lu2008learning}, we use two-member syscall groups by default. Each sequence runs in a separate user-space process, and synchronized release starts both sequences simultaneously.



\section{Concurrent-Input Generation}
\label{sec:input-generation}


\sysname's concurrent-input generation aims to generate diverse syscall groups that expose a broad range of MRPs during concurrent execution, without controlled scheduling. To achieve this goal, in addition to generating individually effective syscall sequences, the fuzzer needs to ensure that the syscall sequences in each group are semantically coordinated to access shared or related kernel state during execution. Therefore, effective syscall-group mutation imposes semantic requirements, including jointly understanding the semantics of syscall sequences in each syscall group, identifying the kernel objects and subsystem states involved, and considering multiple syscall dependencies across the syscall group. 
To satisfy these semantic requirements and improve the diversity of the generated syscall groups, we propose LLM-assisted semantic mutation. By combining its reasoning capabilities with knowledge of system interfaces in prompts, including syscall semantics and kernel-resource lifecycle relations, an LLM can understand the semantics of the syscall group and mutate the two sequences in the group consistently, respecting module-specific syscall, resource, and environment constraints.

\begin{figure}[!t]
\centering
\includegraphics[width=0.98\linewidth]{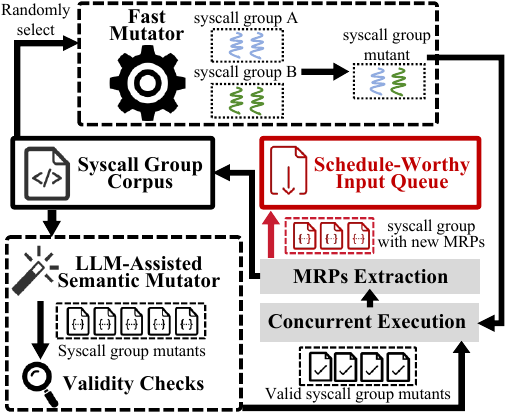}
\caption{Workflow of concurrent-input generation.}
\label{fig:input-workflow}
\end{figure}

\begin{figure}[!t]
\centering
\includegraphics[width=0.98 \linewidth]{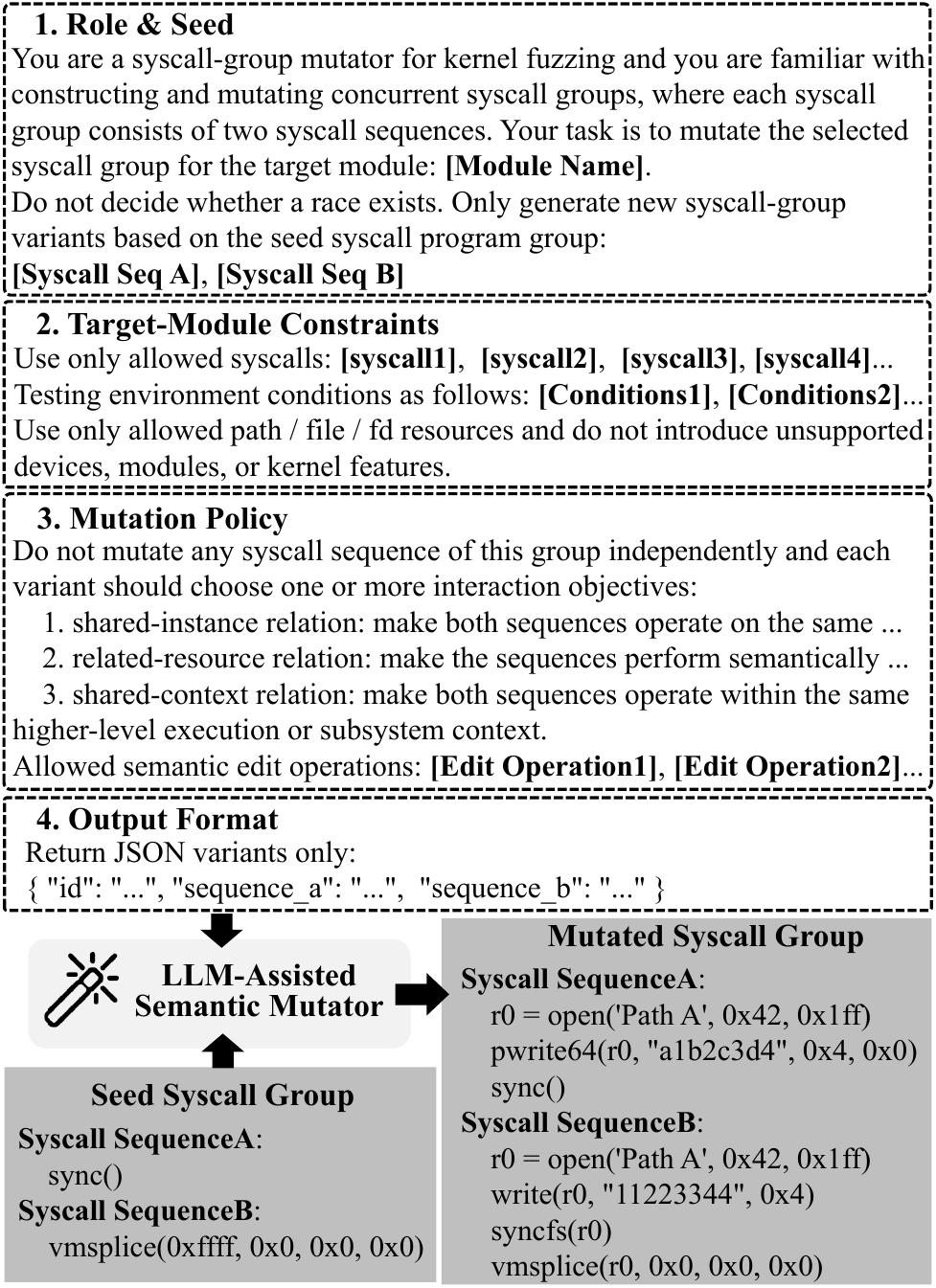}
\caption{LLM-assisted semantic mutation.}
\label{fig:llm-assisted-mutation}
\end{figure}

\subsection{LLM-Assisted Semantic Mutation}
\label{sec:semantic-mutation}

\noindent
\textbf{Workflow.}
Figure~\ref{fig:input-workflow} shows the workflow for concurrent-input generation. \sysname maintains a syscall-group corpus and selects a seed syscall group from the corpus for further mutation. The LLM-assisted semantic mutator jointly rewrites the two syscall sequences in the seed group and generates syscall-group mutants. 
\sysname accepts an LLM mutant only if it has the expected JSON structure and both sequences pass Syzkaller's native deserializer and local checks for enabled syscalls, call limits, allowed paths/devices, and fd dependencies. Accepted groups are then executed concurrently.
Since LLM inference can introduce non-trivial latency, \sysname issues LLM mutation requests asynchronously and submits completed LLM mutants to the validation stage when they become available, without blocking the input-generation pipeline.

While LLM requests are pending, a fast mutator keeps execution supplied by generating and executing syscall groups. When no completed LLM mutant is immediately available, it randomly selects two corpus groups, generates an additional group by randomly pairing sequences sampled from them, and applies Syzkaller's lightweight mutations to each sequence in the syscall group~\cite{syzmutate}.

During concurrent execution, \sysname runs each syscall sequence of a syscall group in a separate user-space process and synchronizes their release, as described in Section~\ref{sec:execution-model}. \sysname then extracts MRPs from the resulting runtime memory-access trace. When an execution contributes one or more new MRPs, the corresponding syscall group is stored in the corpus for further mutation. \sysname also creates a schedule-worthy input entry and adds it to the schedule-worthy input queue for thread scheduling discussed in Section~\ref{sec:thread_scheduling}. 




\vspace{\pspace}
\noindent
\textbf{Mutation.}
 Figure~\ref{fig:llm-assisted-mutation} shows the four-part structured prompt for mutating the selected syscall group. The role \& seed part defines the LLM as a syscall-group mutator and provides the selected seed syscall group. The target-module constraints specify the syscall whitelist, syntax requirements, testing environment, and resource-dependency rules. The LLM receives only these prompt fields; the constraints are distilled offline from the testing environment, kernel source and documentation, and Syzkaller definitions, without sending source code or raw memory-access traces to the LLM. The mutation policy requires the LLM to treat the two sequences as a single mutation unit and coordinate their edits according to predefined cross-sequence interaction objectives. Finally, the output format restricts the response to a structured representation that can be parsed and validated automatically.

The mutation policy defines three complementary cross-sequence interaction objectives. Shared-instance relation specifies that both syscall sequences operate on the same concrete kernel object or resource instance. Related-resource relation coordinates operations on the same resource or on distinct but semantically related resources, such as a directory and one of its files, a link and its target, or different lifecycle stages of a resource. Shared-context relation places both syscall sequences in the same higher-level execution or subsystem context, such as a mount, namespace, device session, protocol state, or subsystem configuration. A generated group may satisfy multiple objectives.

In Figure~\ref{fig:llm-assisted-mutation}, the F2FS example shows how semantic mutation strengthens cross-sequence interaction: the seed loosely combines global synchronization with a file operation, while the mutant adds file writes and synchronization around the same filesystem state, making the two sequences more likely to access overlapping filesystem state through local file operations and global synchronization.


\vspace{\pspace}
\noindent
\textbf{Prioritized seed corpus.}
As in conventional fuzzing, mutation should prioritize interesting seeds. It is especially important for LLM-assisted mutation, as its inference effort for mutating each seed is not trivial. MRP feedback is well-suited to this need because it captures both concrete evidence of cross-thread interaction and the novelty of that interaction. A previously unseen access-site pair represents a new pair of conflicting code locations and thus a new interaction target. For a covered access-site pair, its newly triggered stack contexts indicate a different calling path or kernel state, but its novelty decreases when more stack contexts of that pair have already been observed. 

\section{Thread Scheduling}
\label{sec:thread_scheduling}

\label{sec:thread-interleaving-workflow}
\begin{figure}[!t]
  \centering
  \includegraphics[width=0.98\linewidth]{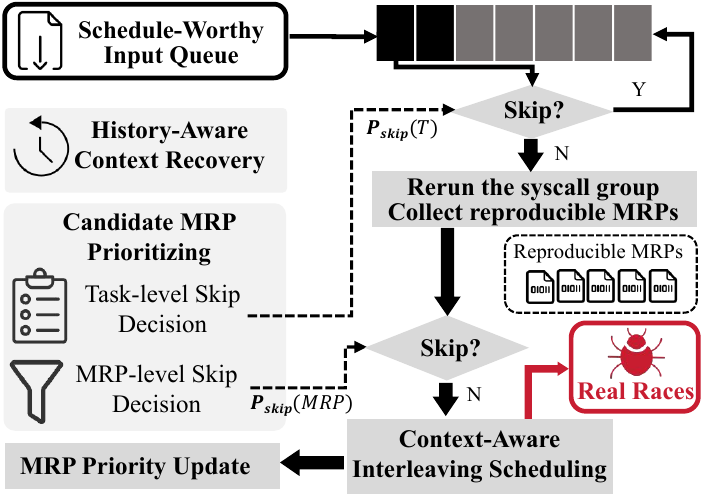}
  \caption{Workflow of thread scheduling.}
  \label{fig:thread-interleaving_workflow}
\end{figure}

Rather than blindly exploring the thread-interleaving space of all generated inputs, \sysname targets the conflicting access pairs specified by schedule-worthy groups' MRPs and schedules threads to force these accesses to occur concurrently, thereby confirming real data races.

As shown in Figure~\ref{fig:thread-interleaving_workflow}, we integrate three techniques to facilitate our scheduling: (1) candidate MRP prioritizing selects more promising syscall groups by excluding syscall groups with confirmed MRPs from further scheduling and deferring syscall groups with repeated unsuccessful scheduling attempts; (2) history-aware context recovery improves the reproducibility of conflicting access pairs by replaying all executed syscalls in a bounded history to reconstruct the required kernel resources, object relationships, and subsystem states; (3) to further improve reproducibility, context-aware interleaving scheduling involves fine-grained thread controls according to the access site, stack context, and the hit counts of specific access sites. 

\sysname dequeues a schedule-worthy entry and applies task-level candidate-MRP prioritization. A skipped task is deferred; otherwise, \sysname replays its bounded history H and reruns syscall group G to reproduce the entry's MRPs. After MRP-level prioritization, the remaining MRPs undergo context-aware interleaving scheduling, which uses the reproduced runtime context to locate target access occurrences, applies a local delay, and updates MRP priority according to the result for subsequent attempts.

\subsection{Candidate MRP Prioritizing}
\label{sec:prioritizing}

During scheduling, each entry dequeued from the schedule-worthy input queue is represented as a scheduling task \(T=\langle H,G,P_G\rangle\), where \(G\) is the schedule-worthy syscall group, \(H\) is its bounded history consisting of previously executed syscalls, and \(P_G\) is the set of MRPs associated with that entry. These MRPs serve as candidate scheduling targets: they indicate conflicting access pairs that may become real data races under targeted interleaving, but have not necessarily been confirmed yet. Candidate MRP prioritizing controls how \sysname spends its limited scheduling budget. 
\sysname updates each candidate MRP's skip probability from prior scheduling results. A higher probability makes it less aggressively attempted under the current budget; if skipped, the candidate is deferred for later scheduling rather than removed.

Candidate MRP prioritizing is applied at two levels. 
At the MRP level,
for each candidate MRP \(p\), \sysname maintains a scheduling state that records whether \(p\) has been confirmed and the number \(F_p\) of failed targeted-scheduling attempts. A candidate with no scheduling state receives \(\Pr_{\mathrm{skip}}(p)=0\), so every newly discovered candidate receives an initial scheduling opportunity. An MRP confirmed as a real race receives \(\Pr_{\mathrm{skip}}(p)=1\), so \sysname avoids redundant scheduling. For the remaining candidates, \sysname computes the skip probability as
\[
\Pr\nolimits_{\mathrm{skip}}(p)=
\min\left(
P_{\max},
\frac{F_p+\alpha}{F_p+\alpha+\beta}\cdot(1-\epsilon)
\right),
\]
where \(\alpha\) and \(\beta\) are smoothing constants, \(\epsilon\) preserves a minimum scheduling probability, and \(P_{\max}\) caps the maximum skip probability allowed for each candidate MRP. 

At the task level, \sysname uses the skip probabilities of the unresolved MRPs in \(P_G\) to decide whether to process a scheduling task \(T=\langle H,G,P_G\rangle\). Let \(U_G \subseteq P_G\) denote the set of candidate MRPs that have not yet been confirmed. \sysname computes the task-level skip probability as the average of their individual skip probabilities:
\[
\Pr\nolimits_{\mathrm{skip}}(T)=
\frac{1}{|U_G|}
\sum_{p_i\in U_G}\Pr\nolimits_{\mathrm{skip}}(p_i).
\]
MRPs confirmed as real races are excluded from \(U_G\). If \(U_G=\emptyset\), the entry has no remaining scheduling target and is retired from the schedule-worthy input queue. Otherwise, \sysname performs a probabilistic skip decision using \(\Pr_{\mathrm{skip}}(T)\). If the entry is skipped, it is appended to the tail of the schedule-worthy input queue rather than discarded. In this paper, we set $\alpha = 0.5$, $\beta = 0.5$, $\epsilon = 0.05$, and $P_{\max} = 0.90$ for candidate MRP prioritization.









\subsection{History-Aware Context Recovery}
\label{sec:history_recovery}

Reproducing candidate MRPs may depend on kernel state created before the syscall group $G$ itself. Such state includes resources, object references, filesystem state, device state, or subsystem state needed to reach the target variable access paths. Therefore, directly executing $G$ from a clean state may fail to reproduce the MRPs in $P_G$. \sysname addresses this problem by replaying a bounded execution history $H$ before executing $G$. Formally, for a triggering syscall group $G_t=G$,
\[
H =
\left\langle
G_{t-N_H},
G_{t-N_H+1},
\ldots,
G_{t-1}
\right\rangle
\]
contains the $N_H$ syscall groups executed immediately before $G_t$, in their original execution order. Replaying only these preceding groups avoids the cost and possible state perturbation of replaying the complete fuzzing history. The configurable bound $N_H$ allows \sysname to balance context recovery effectiveness and replay cost.



Thread scheduling uses this recovery step in both the collection run and the targeted scheduling run. Each run starts from a clean validation VM state, replays \(H\), and then executes \(G\), preventing different scheduling attempts from interfering with each other. The collection run executes \(G\) without targeted delay and records, for each reproduced MRP, the minimum-gap runtime pair that matches the recorded access-site pair and stack-context pair in either access order within \(\tau_{\max}\). The targeted scheduling run executes \(G\) with a targeted delay based on this metadata. Starting each run from a clean VM state would require repeatedly resetting the validation VM. Section~\ref{sec:implementation} describes how \sysname accelerates this process with VM snapshots.

\subsection{Context-Aware Interleaving Scheduling}

Context-aware interleaving scheduling is designed based on the observation that
the same access site may be executed many times under the same stack context even within a single syscall execution. Delaying the wrong accesses of the access sites can miss the target MRP, while delaying all matching occurrences may perturb the original interleaving and drive the execution away from the target race. Therefore, \sysname uses finer-grained runtime context to distinguish different dynamic access occurrences of the same static access site and to locate the two critical occurrences corresponding to the MRP during targeted scheduling. Because concurrent randomness can shift occurrences across different replays, \sysname first uses strict context matching and then progressively relaxes it via fallback strategies rather than relying on a single exact match.


Let \(P_G^r \subseteq P_G\) denote the set of MRPs reproduced during the collection phase. An access sequence number (ASN) is a per-thread counter that orders variable accesses within the same thread, allowing \sysname to distinguish repeated dynamic occurrences of the same static access site. In the preceding collection phase, \sysname selects, for each reproduced MRP, the matching runtime pair with the minimum temporal gap and records its access-site, stack-context, ASN and temporal-gap information. 


Algorithm~\ref{alg:context-aware-scheduling} shows the context-aware scheduling procedure. For each reproduced MRP, \sysname first applies candidate MRP prioritizing at the MRP level. If skipped, the MRP is deferred for future attempts. Otherwise, \sysname uses collection metadata to locate the two target access occurrences. It tries three matching strategies in order. ExactASN matches each access using its access site, stack context, and the exact ASN recorded during collection. If exact matching fails, BoundedASN retains the access-site and stack-context constraints while allowing the ASN to drift within a configured window around the recorded ASN. The window size is configuration-fixed. If both ExactASN and BoundedASN fail, StackOnly drops the ASN constraint and applies the scheduling action to every occurrence matching the static access site and stack context.

After locating the two target accesses, \sysname orders them according to their collection timestamps and denotes the earlier access as \(e_1\). When an occurrence matching \(e_1\) is reached during the targeted scheduling run, \sysname arms a watchpoint-based observation window on the overlapping memory range and delays \(e_1\) by \(\lambda\Delta\), where \(\Delta\) is the temporal gap recorded in the collection phase and \(\lambda\) is a configurable delay multiplier. If the opposite access performs a conflicting access to the overlapping range while this observation window remains active, \sysname confirms the MRP as a real data race. Otherwise, the targeted attempt fails and updates the MRP priority.

\begin{algorithm}[t]
\small
\caption{Context-Aware Interleaving Scheduling}
\label{alg:context-aware-scheduling}
\KwIn{validation task $T=\langle H,G,P_G\rangle$,
re-observed candidate set $P_G^r$ with collection metadata}
\KwOut{confirmed real race set $R$}

$R \leftarrow \emptyset$\;

\ForEach{$p \in P_G^r$}{
    \If{\textsc{SkipMRP}$(p)$}{
        \textsc{Defer}$(p)$\;
        \textbf{continue}\;
    }

    $(e_1,e_2) \leftarrow
    \textsc{OrderedAccesses}(p)$\;

    $\Delta \leftarrow
    \textsc{TimeGap}(p)$\;

    $\mathcal{M} \leftarrow
    [\textsf{ExactASN},
     \textsf{BoundedASN},
     \textsf{StackOnly}]$\;

    \ForEach{$m \in \mathcal{M}$}{
        $\pi \leftarrow
        \textsc{BuildSchedulePlan}
        (e_1,e_2,m,\lambda\Delta)$\;

        $res \leftarrow
        \textsc{ReplayAndSchedule}(H,G,\pi)$\;

        \If{\textsc{ConfirmRace}$(res,p)$}{
            $R \leftarrow R \cup \{(p,m,res)\}$\;
            \textsc{UpdatePriority}$(p,res)$\;
            \textbf{break}\;
        }

        \If{\textsc{TargetMiss}$(res,p)$}{
            \textbf{continue}\;
        }

        \textbf{break}\;
    }

    \If{\textsc{NotConfirmed}$(p)$}{
        \textsc{UpdatePriority}$(p,res)$\;
    }
}

\Return{$R$}\;
\end{algorithm}

\section{Implementation}
\label{sec:implementation}

\sysname extends Syzkaller and adds LLVM-based kernel instrumentation and runtime support.

\noindent
\textbf{Syzkaller extensions.}
We extend Syzkaller's executor to extract MRPs and its fuzzer and manager to generate and mutate syscall groups, manage the MRP corpus and scheduling queue, and control the dynamic threshold. We also add a dedicated validation pipeline for replay and targeted scheduling. These changes comprise about 14.1 KLoC of Go/C++ code. We further add about 1.0 KLoC of Go code for snapshot-based resets in Syzkaller's QEMU interface. QEMU runs the test VMs and accelerates validation resets but does not itself trace memory accesses.

\noindent
\textbf{Kernel instrumentation and runtime support.}
In addition, we implement an LLVM pass and kernel runtime callbacks for memory tracing and validation. The pass instruments memory accesses in selected target modules, and the callbacks emit access records through ftrace for collection by the executor. The callbacks also support the watchpoint-based race confirmation described in Section~\ref{sec:thread_scheduling}. MRP extraction operates directly on the recorded address ranges, as defined in Section~\ref{sec:mrp}, and does not reconstruct high-level object types. This instrumentation and runtime support comprises about 3.3 KLoC of C/C++ code in total.

All reported code sizes are measured with cloc and exclude generated and third-party code.

\section{Evaluation}
\label{sec:evaluation}

\subsection{Experiment Setup}
We evaluate \sysname on eight common Linux kernel modules in Linux 6.17, including four file systems, three device-driver subsystems, and one network protocol stack. We select these modules to cover different parts of the kernel while supporting reproducible testing in a uniform QEMU/Syzkaller environment, allowing us to evaluate the effectiveness of our work across diverse kernel components. The information of these modules is listed in Table~\ref{tab:subjects}. We use CLOC~\cite{cloc} to count their lines of source code.


Experiments run on an AMD Ryzen Threadripper PRO 5975WX 32-core server with 128 GB RAM. We use \textbf{cset} to pin each run to dedicated physical cores to prevent interference. Unless stated otherwise, \sysname splits them evenly between its input-generation and thread-scheduling instances. Each experiment reports its allocated physical cores, memory budget, and run duration.

\begin{table}[!t]
  \centering
  \caption{Information of the tested Linux kernel modules.}
  \label{tab:subjects}
  \footnotesize
  \setlength{\tabcolsep}{2.8pt}
  \renewcommand{\arraystretch}{1}
  \begin{tabularx}{\columnwidth}{lXcc}
    \toprule
    \textbf{Module} & \textbf{Description} & \textbf{Version} & \textbf{LOC} \\
    \midrule
    XFS       & XFS file system                         & Linux 6.17 & 150K \\
    Btrfs     & Btrfs file system                       & Linux 6.17 & 112K \\
    PTMX      & PTMX/TTY subsystem                      & Linux 6.17 & 134K \\
    OSS DSP   & Sound HDA/core subsystem                & Linux 6.17 & 104K \\
    Bluetooth & Bluetooth protocol stack                & Linux 6.17 & 50K  \\
    F2FS      & Flash-Friendly File System              & Linux 6.17 & 39K  \\
    JFS       & Journaled File System                   & Linux 6.17 & 17K  \\
    Floppy    & Floppy block device driver              & Linux 6.17 & 4K   \\
    \bottomrule
  \end{tabularx}
\end{table}

\subsection{Finding Real-World Data Races}


During a three-week continuous testing campaign using DeepSeek-V4-Pro, \sysname found 187 unique real-world data races after deduplication. Based on our manual analysis of their source code and triggering executions, we classify 59 races as harmful and the remaining 128 as benign. We reported all 59 harmful races to the corresponding kernel maintainers. At the time of writing, 39 have been confirmed by the maintainers, of which 15 have been fixed. 

\begin{table}[!t]
  \centering
  \caption{Security-impact classification of the data races.}
  \label{tab:race-impact}
  \footnotesize
  \setlength{\tabcolsep}{2.8pt}
  \renewcommand{\arraystretch}{1}
  \begin{tabular*}{\columnwidth}{@{\extracolsep{\fill}}lcccc@{}}
    \toprule
    Module &
    \shortstack{Harmful\\ / Benign} &
    \shortstack{Memory Corruption\\ / DoS} &
    \shortstack{Data\\ Corruption} &
    \shortstack{Unexpected\\ Behavior} \\
    \midrule
    XFS       & 5 / 17   & 1 & 4  & 0 \\
    Btrfs     & 2 / 19   & 0 & 2  & 0 \\
    PTMX      & 6 / 14   & 0 & 6  & 0 \\
    OSS DSP   & 11 / 2   & 0 & 5  & 6 \\
    Bluetooth & 7 / 2    & 5 & 2  & 0 \\
    F2FS      & 8 / 33   & 1 & 6  & 1 \\
    JFS       & 11 / 19  & 0 & 10 & 1 \\
    Floppy    & 9 / 22   & 0 & 4  & 5 \\
    \midrule
    Total    & 59 / 128 & 7 & 39 & 13 \\
    \bottomrule
  \end{tabular*}
\end{table}

\subsection{Security Impact Analysis}
\label{sec:security-impact-analysis}

Table~\ref{tab:race-impact} summarizes the security-impact classification of the detected data races. Among the 59 harmful races, 7 can cause memory corruption (\emph{e.g.}, use-after-frees) or denial of service (DoS).
39 data races can corrupt critical kernel state or persistent data. Specifically, they
change critical variables used in kernel state management, and the raced values
can lead to inconsistent updates of file system state, device state, request state, or persistent metadata. 
13 data races can cause unexpected behavior by affecting critical control-flow or state-machine decisions in the kernel modules.
We present one representative example from each category.

\begin{figure}[t]
  \centering
  \includegraphics[width=0.98\linewidth]{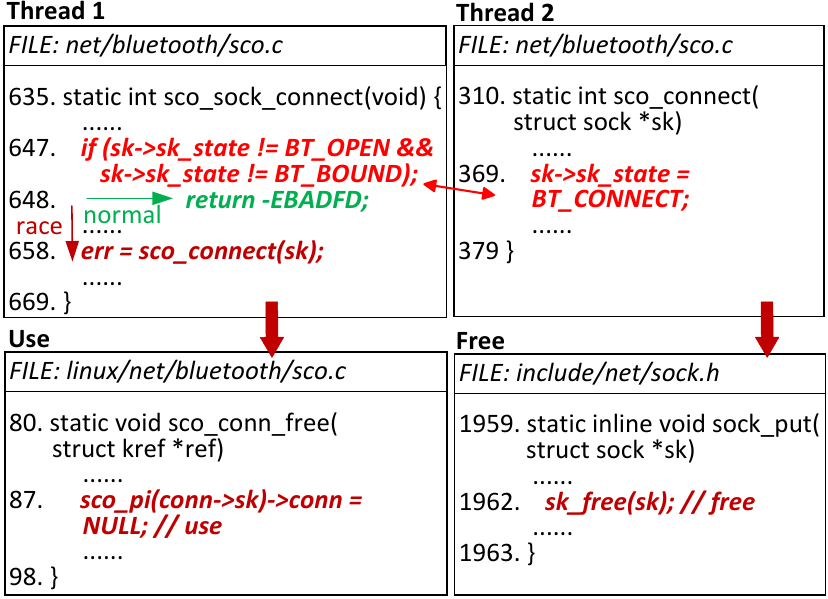}
  \caption{A harmful data race found in the Bluetooth stack.}
  \label{fig:bt-mem-corruption-dos}
\end{figure}

\vspace{\pspace}
\noindent
\textbf{Memory corruption.}
Figure~\ref{fig:bt-mem-corruption-dos} shows a harmful race in the Bluetooth SCO connect path. Here, \texttt{BT\_OPEN}, \texttt{BT\_BOUND}, and \texttt{BT\_CONNECT} are Bluetooth socket states. Thread~1 checks the SCO socket state without holding the socket lock, while Thread~2 can concurrently update the same state in \texttt{sco\_connect()}. The read in \texttt{sco\_sock\_connect()} and the write in \texttt{sco\_connect()} therefore form a data race. In this case, Thread~1 can pass the check with the state value \texttt{BT\_OPEN} or \texttt{BT\_BOUND}, and then Thread~2 assigns \texttt{BT\_CONNECT} to the socket state.
In the end, the two threads operate on the same socket with inconsistent connection ownership. As a result, the HCI disconnect handler for the connection installed by the earlier connect attempts may free the socket reference, while the later concurrent connect attempts in the other thread still access the socket through a stale SCO connection object whose associated socket has already been freed. This data race, therefore, triggers a use-after-free in the Bluetooth subsystem, which can corrupt kernel memory or crash the whole kernel.



\begin{figure}[t]
  \centering
  \includegraphics[width=0.98\linewidth]{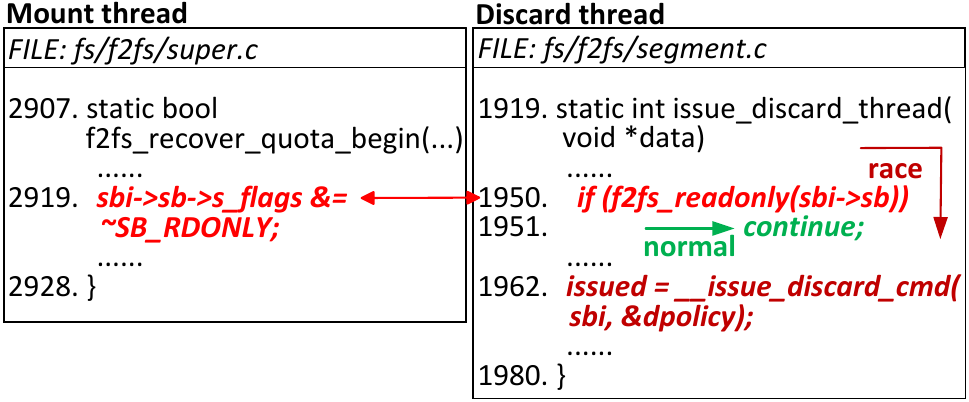}
  \caption{A harmful data race found in F2FS.}
  \label{fig:F2FS-datacorruption}
\end{figure}


\vspace{\pspace}
\noindent
\textbf{Data corruption.} Figure~\ref{fig:F2FS-datacorruption} shows a harmful race in F2FS. During recovery of a read-only mount, the mount thread temporarily clears a flag \texttt{SB\_RDONLY} to update recovery metadata, while the discard thread 
checks the same flag to decide whether discard I/O is allowed. 
These two accesses can occur concurrently during recovery and result in a data race. If the mount thread clears the flag first, the discard thread will observe the temporary flag value, which is not \texttt{SB\_RDONLY}. In this case, the discard thread issues the discard or zone-reset requests before the mount thread restores the read-only flag. Such a data race makes a recovery-only state transition escape to the block device and discard ranges that still contain file-system data or metadata. As a result, it causes persistent data corruption.


\begin{figure}[!t]
  \centering
  \includegraphics[width=0.98\linewidth]{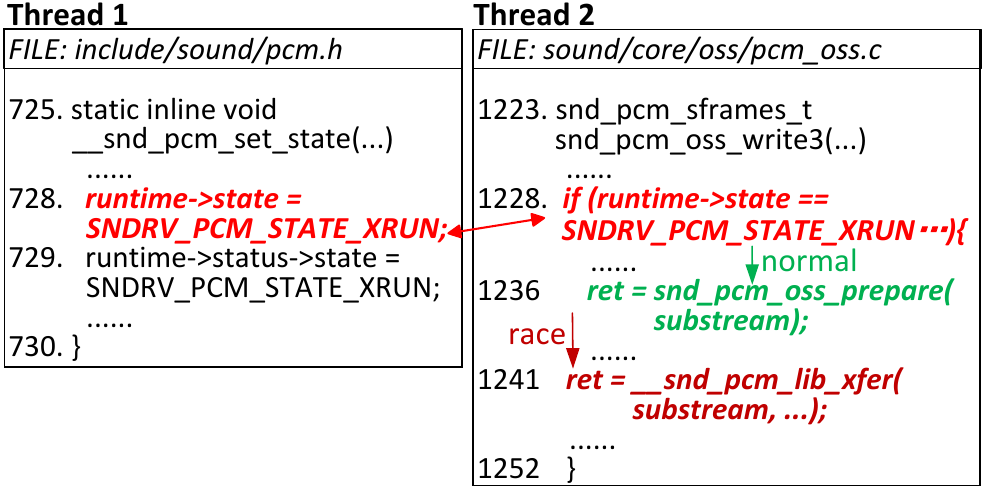}
  \caption{A harmful data race found in ALSA.}
  \label{fig:alsa-undefined-behavior}
\end{figure}


\vspace{\pspace}
\noindent
\textbf{Unexpected behavior.} Figure~\ref{fig:alsa-undefined-behavior} shows a data race in ALSA OSS PCM playback. In Thread~1, the IRQ-driven XRUN handling code updates \texttt{runtime->state} to \texttt{XRUN}. In Thread~2, the OSS write code reads \texttt{runtime->state} to decide whether recovery is required before transferring samples. 
Such two accesses can occur concurrently and cause a data race.
If Thread~2 reads a stale \texttt{RUNNING} value from \texttt{runtime->state} before \texttt{runtime->state} is assigned \texttt{XRUN} in Thread 1, Thread 2 can skip the required XRUN recovery and enter the transfer path while the stream is in the XRUN state. Consequently, the write operation proceeds without recovering the stream and fails instead of successfully restoring normal playback.

\subsection{Ablation Study on Dynamic Threshold Control}
\label{sec:dynamic-threshold-ablation}

\vspace{\pspace}
\noindent
\textbf{Experiment setup.}
We conduct a 24-hour ablation on all eight modules using seven otherwise identical configurations.
Dynamic follows Algorithm~1; Fixed-min and Fixed-max keep $\tau$ at $\tau_{\min}$ and $\tau_{\max}$;
Fixed-25\%, Fixed-50\%, and Fixed-75\% keep $\tau$ at
$\tau_{\min}+q(\tau_{\max}-\tau_{\min})$, where $q=0.25,0.5,0.75$, respectively; and Random-$\tau$ samples $\tau$ uniformly from the same interval at each control interval.


Dynamic confirms 76 data races, compared with 39 for Random-$\tau$ and 12, 55, 28, 24, and 28 for Fixed-min, Fixed-25\%, Fixed-50\%, Fixed-75\%, and Fixed-max, respectively. It performs best on six evaluated modules and ties the best fixed setting on Floppy and Bluetooth.
These results demonstrate that the effectiveness of our dynamic threshold control: by responding to queue backpressure, it mitigates scheduler undersupply at small thresholds and wasted scheduling effort on low-potential candidates at large thresholds, thereby improving data-race detection over the tested fixed and randomly sampled thresholds.

\begin{table}[t]
    \centering
    \caption{Confirmed data races under threshold policies.}
    \label{tab:dynamic-threshold-ablation}
    \footnotesize
    \setlength{\tabcolsep}{1.5pt}
    \renewcommand{\arraystretch}{1}
    \begin{tabular*}{\columnwidth}{
      @{\extracolsep{\fill}}lrrrrrrrrr@{}
    }
      \toprule
      \textbf{Setting}
        & \textbf{XFS}
        & \textbf{F2FS}
        & \textbf{Btrfs}
        & \textbf{PTMX}
        & \textbf{JFS}
        & \textbf{Floppy}
        & \textbf{DSP}
        & \textbf{Bluetooth}
        & \textbf{Total} \\
      \midrule
      Dynamic
        & 6 & 22 & 11 & 13 & 6 & 5 & 4 & 9 & 76 \\
      Random-\(\tau\)
        & 3 & 7 & 6 & 11 & 0 & 4 & 1 & 7 & 39 \\
      Fixed-min
        & 1 & 3 & 1 & 3 & 1 & 2 & 0 & 1 & 12 \\
      Fixed-25\%
        & 4 & 16 & 10 & 12 & 2 & 5 & 1 & 5 & 55 \\
      Fixed-50\%
        & 1 & 9 & 6 & 4 & 0 & 4 & 0 & 4 & 28 \\
      Fixed-75\%
        & 0 & 7 & 2 & 5 & 0 & 3 & 0 & 7 & 24 \\
      Fixed-max
        & 2 & 6 & 2 & 7 & 0 & 2 & 0 & 9 & 28 \\
      \bottomrule
    \end{tabular*}
\end{table}
\subsection{Ablation Study on Concurrent-Input Mutation}
\label{sec:input-generation-analysis}


\begin{figure*}[t]
  \centering
  \includegraphics[width=0.99\textwidth]{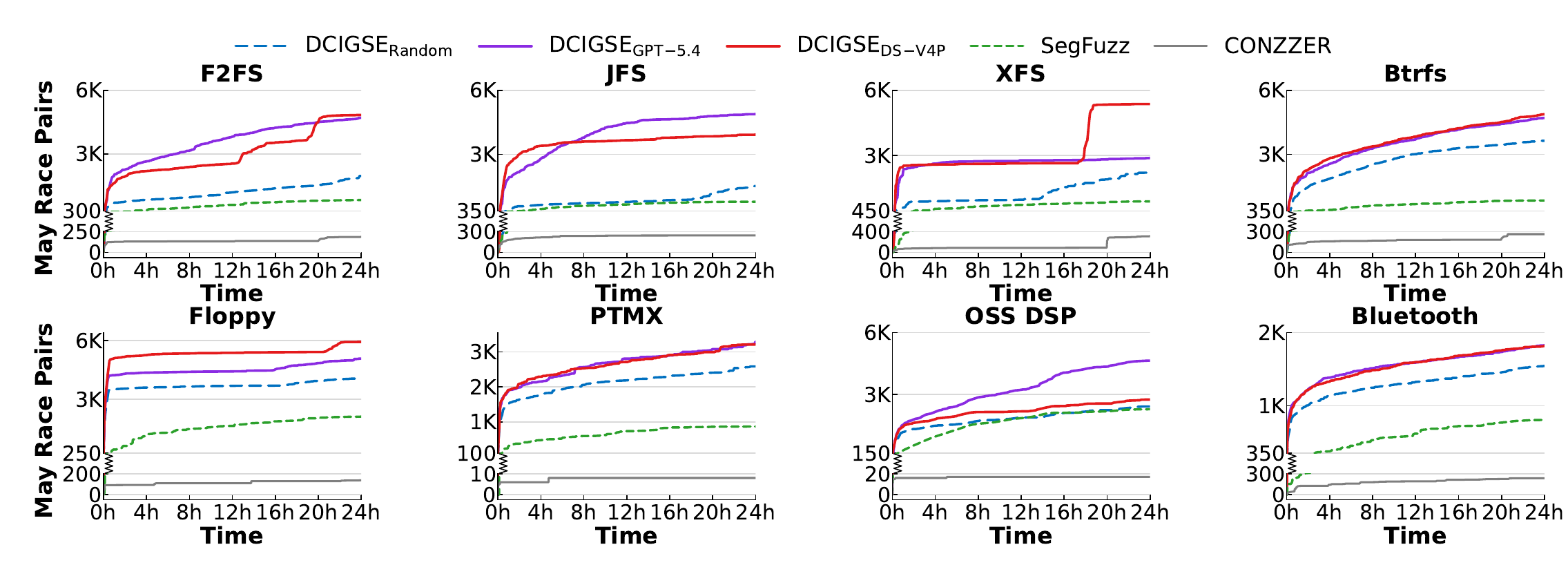}
  \caption{Growth of MRPs produced by different \sysname mutation strategies and prior kernel concurrency fuzzers.}
  \label{fig:mrp_number}
\end{figure*}

\noindent
\textbf{Experiment setup.}
This experiment evaluates whether LLM-assisted semantic mutation improves concurrent-input generation and increases data race detection effectiveness. We compare two LLMs, GPT-5.4 and DeepSeek-V4-Pro, with random mutation. The random mutation baseline involves no LLM and uses only our fast mutator described in Section~\ref{sec:input-generation}, which samples syscall sequences from the same syscall-group corpus, randomly mutates each sequence with syz-mutate, and then forms syscall groups. The three configurations use the same complete \sysname framework and differ only in their input mutation strategy. For each kernel module, we run each configuration for 24 hours using four dedicated physical CPU cores and a 16 GB memory budget per run. In this experiment, we record the growth of MRPs collected under each configuration. 



\vspace{\pspace}
\noindent
\textbf{Analysis on LLM-assisted semantic mutation.}
Figure~\ref{fig:mrp_number} reports the resulting MRP growth curves. Compared with the no-LLM Random configuration, which uses only our fast mutator while keeping all other components unchanged, both LLM-assisted configurations generally produce more MRPs, especially on file systems. This comparison isolates the contribution of LLM-assisted semantic mutation: it generates syscall groups whose syscalls are more likely to access shared or related kernel state and trigger MRPs.

More MRPs are useful only if they result in more confirmed races. Table~\ref{tab:input-validate-races} shows that the GPT-5.4 and DeepSeek-V4-Pro configurations confirm 76 and 79 races, respectively, compared with 41 for random mutation. This corresponds to improvements of 85.4\% and 92.7\%. Both LLM-assisted configurations outperform random mutation on each evaluated kernel module. The similar gains obtained with two different LLMs suggest that our semantic mutation can work well with different LLMs in practice.

\begin{table}[t]
    \centering
    \caption{Confirmed data races under different input-mutation strategies.}
    \label{tab:input-validate-races}
    \footnotesize
    \setlength{\tabcolsep}{1.5pt}
    \renewcommand{\arraystretch}{1}
    \begin{tabular*}{\columnwidth}{
      @{\extracolsep{\fill}}lrrrrrrrrr@{}
    }
      \toprule
      \textbf{Setting}
        & \textbf{XFS}
        & \textbf{F2FS}
        & \textbf{Btrfs}
        & \textbf{PTMX}
        & \textbf{JFS}
        & \textbf{Floppy}
        & \textbf{DSP}
        & \textbf{Bluetooth}
        & \textbf{Total} \\
      \midrule
      GPT-5.4
        & 6 & 22 & 11 & 13 & 6 & 5 & 4 & 9 & 76 \\
      DS-V4P
        & 8 & 17 & 11 & 16 & 5 & 10 & 5 & 7 & 79 \\
      Random
        & 4 & 7 & 9 & 8 & 3 & 3 & 2 & 5 & 41 \\
      \bottomrule
    \end{tabular*}
\end{table}


  \begin{table*}[t]
    \centering
    \caption{GPT-5.4 API cost and token usage under dynamic threshold control. Parentheses denote token counts in millions.}
    \label{tab:llm-cost}
    \footnotesize
    \setlength{\tabcolsep}{2.4pt}
    \renewcommand{\arraystretch}{1}
    \begin{tabular*}{\textwidth}{@{\extracolsep{\fill}}lcccccccc@{}}
      \toprule
      \textbf{Type} &
      \textbf{XFS} & \textbf{F2FS} & \textbf{Btrfs} &
      \textbf{PTMX} & \textbf{JFS} & \textbf{Floppy} &
      \textbf{DSP} & \textbf{Bluetooth} \\
      \midrule
      Uncached input
        & \$0.41 (0.16M) & \$1.89 (0.76M) & \$4.50 (1.80M)
        & \$2.63 (1.05M) & \$0.31 (0.12M) & \$1.21 (0.48M)
        & \$2.89 (1.16M) & \$3.28 (1.31M) \\
      Cached input
        & \$0.02 (0.07M) & \$0.02 (0.09M) & \$0.21 (0.85M)
        & \$0.18 (0.72M) & \$0.04 (0.14M) & \$0.09 (0.37M)
        & \$0.22 (0.88M) & \$0.21 (0.84M) \\
      Output
        & \$0.92 (0.06M) & \$2.58 (0.17M) & \$8.65 (0.58M)
        & \$8.74 (0.58M) & \$1.04 (0.07M) & \$4.11 (0.27M)
        & \$13.55 (0.90M) & \$8.29 (0.55M) \\
      \midrule
      Total
        & \$1.35 (0.29M) & \$4.49 (1.02M) & \$13.37 (3.23M)
        & \$11.55 (2.36M) & \$1.39 (0.34M) & \$5.41 (1.13M)
        & \$16.67 (2.94M) & \$11.78 (2.70M) \\
      \bottomrule
    \end{tabular*}
  \end{table*}
  

To quantify the cost of LLM-assisted mutation, we recorded GPT-5.4 API usage. Table~\ref{tab:llm-cost} separates uncached input, cached input, and output tokens. Each entry gives USD followed by millions of tokens in parentheses. Costs use the official standard rates of \$2.50, \$0.25, and \$15.00 per million tokens. Across the eight runs, usage totaled 14.01M tokens and \$66.01. Inference used a remote API and required no local GPU hardware.
  

\subsection{Ablation Study on Thread Scheduling}

\noindent
\textbf{Experiment setup.}
This experiment evaluates the contributions of each design component in the thread-scheduling instance of \sysname. Unlike concurrent-input generation, thread scheduling does not aim to generate more May Race Pairs. Instead, it aims to efficiently schedule the thread execution of the target syscall groups, forcing the conflicting access pair specified in the MRP to occur concurrently.


To isolate the effects of thread scheduling from randomness in concurrent-input generation, all experiments in this section use a corpus of syscall groups with MRPs for each module. The corpus is generated by the DeepSeek-V4-Pro configuration described in Section~\ref{sec:input-generation-analysis}. For each module, all thread scheduling configurations consume the same corpus and use the same scheduling budget, differing only in the enabled design components. For each kernel module, we run each configuration for 24 hours using two dedicated physical CPU cores and an 8~GB memory budget.


We compare the full configuration with three ablated configurations. \textsc{No-history} disables history-aware context recovery and executes each candidate group without replaying its saved execution history. \textsc{No-ASN} disables ASN-based occurrence matching and uses only \textsc{StackOnly}, retaining access-site and stack-context constraints but applying delay to every matching dynamic occurrence. \textsc{No-backoff} disables probabilistic candidate skipping and deferral based on previous scheduling outcomes.

  \begin{table}[!t]
    \centering
    \caption{Effect of thread-scheduling components on confirmed data races.}
    \label{tab:validate-component-sensitivity}

    \footnotesize
    \setlength{\tabcolsep}{1pt}
    \renewcommand{\arraystretch}{1}

    \begin{tabular*}{\columnwidth}{
      @{\extracolsep{\fill}}lccccccccc@{}
    }
      \toprule
      \textbf{Setting}
        & \textbf{XFS}
        & \textbf{Btrfs}
        & \textbf{PTMX}
        & \textbf{DSP}
        & \textbf{Bluetooth}
        & \textbf{F2FS}
        & \textbf{JFS}
        & \textbf{Floppy}
        & \textbf{Total} \\
      \midrule
      Full
        & 7 & 12 & 18 & 6 & 6 & 18 & 5 & 11 & 83 \\
      No-history
        & 2 & 5 & 8 & 3 & 5 & 8 & 1 & 3 & 35 \\
      No-ASN
        & 2 & 7 & 9 & 5 & 4 & 11 & 2 & 6 & 46 \\
      No-backoff
        & 3 & 5 & 13 & 0 & 4 & 7 & 2 & 6 & 40 \\
      \bottomrule
    \end{tabular*}

  \end{table}




\begin{table*}[t]
\centering
\caption{Scheduling-budget utilization by module.}
\label{tab:validation-budget-utilization-module}

\footnotesize
\setlength{\tabcolsep}{3pt}
\renewcommand{\arraystretch}{1}

\begin{tabular*}{\textwidth}{
  @{\extracolsep{\fill}}lrrrrrrrrr@{}
}
  \toprule
  \textbf{Metric}
    & \textbf{XFS}
    & \textbf{Btrfs}
    & \textbf{PTMX}
    & \textbf{OSS DSP}
    & \textbf{Bluetooth}
    & \textbf{F2FS}
    & \textbf{JFS}
    & \textbf{Floppy}
    & \textbf{Total} \\
  \midrule
  Full skips
    & 297 & 1,549 & 462 & 706 & 220 & 2,264 & 463 & 2,263
    & \textbf{8,224} \\
  Full processed
    & 712 & 2,103 & 1,013 & 971 & 450 & 3,064 & 950 & 3,222
    & \textbf{12,485} \\
  No-ASN skips
    & 511 & 1,528 & 427 & 835 & 202 & 2,225 & 511 & 2,515
    & \textbf{8,754} \\
  No-ASN processed
    & 1,008 & 2,146 & 979 & 1,123 & 450 & 3,046 & 1,001 & 3,565
    & \textbf{13,318} \\
  No-backoff processed
    & 407 & 504 & 546 & 276 & 236 & 774 & 489 & 967
    & \textbf{4,199} \\
  \bottomrule
\end{tabular*}
\end{table*}
  
Tables~\ref{tab:validate-component-sensitivity} and~\ref{tab:validation-budget-utilization-module} report confirmed races and budget utilization. A processed pair is either scheduled or skipped and deferred, so attempts equal processed pairs minus skips.

\vspace{\pspace}
\noindent
\textbf{Analysis on history-aware context recovery.}
As shown in Table~\ref{tab:validate-component-sensitivity}, history-aware context recovery substantially improves candidate reproducibility. The full configuration confirms 83 real data races, whereas No-history confirms only 35 real data races, indicating a reduction of 57.8\%. The effectiveness degradation appears across all evaluated modules. Without recent nearby execution history, a candidate group may fail to reconstruct the file system, device, resource-reference, or protocol state under which its target conflicting access pairs were originally observed. The data races still confirmed by No-history indicate that some syscall groups are largely self-contained, but the large reduction shows that many candidate MRPs depend on state established by earlier executions.

\vspace{\pspace}
\noindent
\textbf{Analysis on context-aware interleaving scheduling.}
Context-aware interleaving scheduling substantially improves target-localization precision. No-ASN confirms 46 real data races, compared with 83 for the full configuration, indicating a reduction of 44.6\%. This effectiveness degradation is not caused by fewer scheduling opportunities. No-ASN executes 4,564 targeted scheduling attempts and processes 13,318 candidate pairs, compared with 4,261 attempts and 12,485 processed pairs for the full configuration. Without ASN information, multiple dynamic occurrences sharing the same access site and stack context may be treated as the target occurrence. Applying delay to these additional occurrences perturbs the execution without reliably bringing the intended MRP endpoints into overlap, reducing the effectiveness of targeted interleaving exploration.

\vspace{\pspace}
\noindent
\textbf{Analysis on candidate MRP prioritizing.}
Candidate MRP prioritizing improves the allocation of the scheduling budget. No-backoff confirms 40 data races, compared with 83 for the full configuration, indicating a reduction of 51.8\%. The two configurations execute a similar number of scheduling attempts: 4,199 for No-backoff and 4,261 for the full configuration. However, the full configuration skips or defers 8,224 low-priority candidates and consequently advances over 12,485 candidate pairs, approximately 2.97 times the 4,199 pairs reached by No-backoff. Without backoff, low-potential candidates continue to consume scheduling attempts and prevent interleaving exploration from reaching a broader portion of the MRP corpus.

Overall, the three mechanisms address different challenges in thread scheduling for confirming data races. History-aware context recovery reconstructs the state required to reproduce candidate accesses, ASN-aware matching anchors scheduling delay to the intended dynamic occurrences, and candidate MRP prioritizing prevents low-potential candidates from repeatedly monopolizing the scheduling budget. Together, these mechanisms enable \sysname to confirm substantially more MRP candidates as real data races with efficient thread scheduling under a bounded scheduling budget.

\subsection{Comparison with Prior Approaches}

\noindent
\textbf{Experiment setup.}
We compare \sysname with two state-of-the-art concurrency-aware kernel fuzzers, SegFuzz~\cite{segfuzz} and CONZZER~\cite{conzzer}. We examine two complementary outcomes from the same 24-hour end-to-end runs: the growth of MRPs during execution and the number of real data races confirmed by the end of the run. SegFuzz and CONZZER were originally evaluated on Linux v6.2 and v5.4, respectively~\cite{segfuzz,conzzer}. We ported both implementations to v6.17 with compatibility changes only, thereby preserving their original functionality.

For \sysname, we reuse the three configurations evaluated in Section~\ref{sec:input-generation-analysis}: $\mathrm{\sysname}_{\mathrm{GPT\text{-}5.4}}$, which uses GPT-5.4-assisted semantic mutation; $\mathrm{\sysname}_{\mathrm{DS\text{-}V4P}}$, which uses DeepSeek-V4-Pro-assisted semantic mutation; and $\mathrm{\sysname}_{\mathrm{Random}}$, which uses random syscall-group mutation. In all three configurations of \sysname, the input-generation and thread-scheduling instances run in parallel, and the dynamic time-threshold controller remains enabled. SegFuzz and CONZZER run in their original modes. All approaches run on Linux v6.17 using the same evaluation setup, \emph{i.e.}, four physical CPU cores, 16 GB memory, and 24-hour testing per kernel module.


We extract MRPs from the executions of SegFuzz and CONZZER using the MRP construction and accounting rules described in Section~\ref{sec:mrp}. Because these baselines do not implement \sysname's dynamic threshold controller, we use \(\tau_{\max}\), the maximum threshold permitted by the controller, when extracting their MRPs. This permissive setting admits conflicting access pairs with larger temporal distances and therefore avoids artificially limiting the number of MRPs attributed to the baselines.

\vspace{\pspace}
\noindent
\textbf{MRP analysis.}
Figure~\ref{fig:mrp_number} shows the growth of MRPs during runs. The \sysname curves are the same results analyzed in Section~\ref{sec:input-generation-analysis} and are included here for direct comparison with SegFuzz and CONZZER. The three \sysname configurations generally produce more MRPs across the evaluated kernel modules, with clear advantages on the file systems. Importantly, $\mathrm{\sysname}_{\mathrm{Random}}$ produces more MRPs than the two baselines. Since this configuration does not use LLM-assisted mutation, this result demonstrates the benefit of decoupling concurrent-input generation and thread-interleaving exploration through the MRP interface, which significantly improves the fuzzing efficiency.

\vspace{\pspace}
\noindent
\textbf{Data race detection.}
Table~\ref{tab:comparison-real-races} reports the confirmed data races from the same end-to-end runs. The \sysname rows repeat Table~\ref{tab:input-validate-races} for direct per-module comparison with the baselines. $\mathrm{\sysname}_{\mathrm{DS\text{-}V4P}}$ confirms 79 data races versus 19 for SegFuzz and 11 for CONZZER, improvements of 4.16\(\times\) and 7.18\(\times\). Even $\mathrm{\sysname}_{\mathrm{Random}}$ confirms 41 data races, 2.16\(\times\) and 3.73\(\times\) those of SegFuzz and CONZZER. This result shows that the improvement does not rely solely on LLM-assisted mutation: decoupling concurrent-input generation and thread-interleaving exploration already provides a substantial empirical advantage over the baselines that couple these two phases. 

These results show that \sysname improves both the supply of concrete scheduling targets (\emph{i.e.}, MRPs) and their conversion into real data races. The decoupled framework provides significant improvement over existing approaches. 

\begin{table}[t]
  \centering
  \caption{Confirmed data races across approaches.}
  \label{tab:comparison-real-races}

  \footnotesize
  \setlength{\tabcolsep}{3.2pt}
  \renewcommand{\arraystretch}{1}

  \begin{tabular*}{\columnwidth}{
    @{\extracolsep{\fill}}
    lccccc
    @{}
  }
    \toprule
    \multirow{2}{*}{\textbf{Module}}
      & \multicolumn{3}{c}{\textbf{\sysname{}}}
      & \multirow{2}{*}{\textbf{SegFuzz}}
      & \multirow{2}{*}{\textbf{CONZZER}} \\
    \cmidrule(lr){2-4}
      & \textbf{GPT-5.4}
      & \textbf{DS-V4P}
      & \textbf{Random}
      & 
      & \\
    \midrule
    XFS
      & 6  & 8  & 4 & 3 & 1 \\
    F2FS
      & 22 & 17 & 7 & 6 & 3 \\
    Btrfs
      & 11 & 11 & 9 & 2 & 1 \\
    PTMX
      & 13 & 16 & 8 & 1 & 1 \\
    JFS
      & 6  & 5  & 3 & 3 & 4 \\
    Floppy
      & 5  & 10 & 3 & 2 & 1 \\
    OSS DSP
      & 4  & 5  & 2 & 1 & 0 \\
    Bluetooth
      & 9  & 7  & 5 & 1 & 0 \\
    \midrule
    \textbf{Total}
      & 76
      & 79
      & 41
      & 19
      & 11 \\
    \bottomrule
  \end{tabular*}

\end{table}

\section{Related Work}

\subsection{Coverage-Guided Fuzzing}



Coverage-guided fuzzing, like AFL, uses execution coverage to collect interesting inputs for further mutation~\cite{afl}. Later work improves this paradigm through better seed scheduling and search strategies~\cite{aflfast,aflgo,aflplusplus}, or through more informative feedback and structure-aware mutation~\cite{angora,fairfuzz,smartgreybox,superion}. These techniques increase code coverage and expose diverse sequential paths. However, such feedback mainly captures sequential paths and cannot reliably indicate an input’s potential to trigger interesting concurrent executions. In contrast, MRPs capture conflicting variable-access pairs that occur within a short time interval, providing direct guidance for focusing computing resources on syscall groups that are more likely to expose data races.

\subsection{Kernel and Concurrency Fuzzing}
\begingroup
\looseness=-1
Coverage-guided fuzzing has been widely applied to kernel testing, where inputs are typically represented as syscall programs. Syzkaller and kAFL establish the basic coverage-guided kernel fuzzing framework~\cite{syzkaller,kafl}. Subsequent systems improve kernel input generation by better modeling syscall semantics, exploring complex kernel states, and supporting diverse kernel interfaces and subsystems~\cite{moonshine,difuze,syzgen,syzdescribe,syzgenplusplus,kernelgpt,janus,hydra,statefuzz,syzdirect,actor,usbfuzz,periscope,printfuzz,charm,agamotto}. These techniques substantially broaden the kernel code paths exercised by generated inputs, but they still focus solely on exploring the input space without considering thread interleavings. As a result, they may miss data races that manifest only under rare or specific interleavings.


Existing kernel concurrency fuzzing additionally explores the thread-interleaving space. Prior work improves this thread scheduling through search space reduction, concurrency-aware feedback, and controlled interleaving scheduling~\cite{chess,ski,datacollider,razzer,krace,conzzer,segfuzz,snowboard,snowcat,rff,sect}. However, most existing approaches still couple concurrent-input generation with interleaving exploration, causing substantial testing effort to be spent scheduling inputs with little potential to trigger data races. In contrast, \sysname decouples these two searches: its input-generation stage collects MRPs during concurrent execution, while its interleaving-exploration stage uses the collected MRPs to concentrate interleaving exploration on promising regions of the interleaving space.
\endgroup

\subsection{LLM-Assisted Testing}
\begingroup
\looseness=-1


LLMs have been increasingly applied to software testing~\cite{codamosa,chatunitest,testgenllm}, and various fuzzing tasks such as input generation, mutation, and fuzzer synthesis~\cite{chatafl,fuzz4all,titanfuzz,ckgfuzzer,elfuzz,g2fuzz,pangolin}. In kernel fuzzing, KernelGPT uses LLMs to synthesize and refine Syzkaller syscall specifications, while SyzAgent uses them to guide test-case generation and mutation during kernel fuzzing~\cite{kernelgpt,syzagent}. Themis uses LLMs to generate RPC-based tests for statically identified distributed race candidates, followed by directed fuzzing of their input parameters~\cite{themis}. KnitFuzz uses LLMs to generate context-aware C seeds for socket-state exploration~\cite{knitfuzz}. In contrast, \sysname semantically mutates concurrent syscall groups while preserving and strengthening their interactions. Although prior LLM-based kernel fuzzers improve syscall specifications or context-aware input generation, they do not make semantic interactions among concurrently executed inputs an explicit mutation objective. Unlike prior LLM-based kernel fuzzers, \sysname explicitly targets concurrency semantics during mutation to explore the concurrent-input space more effectively and improve data-race detection.
\endgroup

\section{Conclusion}
This paper presents \sysname, a kernel concurrency fuzzing framework that decouples concurrent-input generation and thread-interleaving exploration through May Race Pairs (MRPs). By using MRPs to guide semantic mutation during concurrent-input generation and to focus subsequent scheduling effort on promising interleavings, \sysname avoids spending substantial testing effort on inputs with little potential to trigger data races. Our evaluation on eight Linux kernel modules found 187 data races, including 59 harmful data races, 39 of which have been confirmed by kernel maintainers and 15 of which have been fixed. Under the same testing budget, \sysname detects substantially more data races than state-of-the-art kernel concurrency fuzzers.

More broadly, \sysname shows that lightweight data race candidate discovery can be decoupled from expensive thread scheduling through an explicit runtime abstraction. MRP instantiates this abstraction specifically for data races, while other abstractions that are inexpensive to collect and informative enough to guide scheduling can support other concurrency bugs. For example, deallocation/use pairs that record a deallocation in one thread and an access to the same object in another thread can decouple the searches: concurrent-input generation generates inputs with greater potential for concurrency-induced UAFs, while scheduling targets the unsafe order. We believe our decoupling design offers a general and effective paradigm for concurrency fuzzing beyond the specific design of \sysname.



\bibliographystyle{IEEEtran}
\bibliography{references}

\appendices


\newpage 





\end{document}